\documentclass[12pt,a4paper]{article}
\usepackage{amssymb,amsmath}
\usepackage{a4wide}
\usepackage{cite}
\usepackage[hyperindex=true,
          pdfstartview=FitH,
          bookmarksnumbered=true,
          bookmarksopen=true,
          citecolor=blue,
          linkcolor=blue,
          colorlinks=true,
          pdfborder=001,
          unicode]{hyperref}

\allowdisplaybreaks
\numberwithin{equation}{section}

\begin{document}
\title{Entropy relations and the application of black holes with cosmological constant and  Gauss-Bonnet term}
\author{Wei Xu$^{1}$\thanks{{\em
        email}: \href{mailto:xuweifuture@gmail.com}
        {xuweifuture@gmail.com}}\ ,
        Jia Wang$^{2}$\thanks{{\em
        email}: \href{mailto:wangjia2010@mail.nankai.edu.cn}{wangjia2010@mail.nankai.edu.cn}}
         and Xin-he Meng$^{2,3}$\thanks{{\em
        email}: \href{mailto:xhm@nankai.edu.cn}
        {xhm@nankai.edu.cn}}\\
$^{1}$School of Physics, Huazhong University of Science and Technology, \\ Wuhan 430074, China\\
$^{2}$School of Physics, Nankai University, Tianjin 300071, China\\
$^{3}$ State Key Laboratory of ITP, ITP-CAS, Beijing 100190, China}
\date{}
\maketitle
\begin{abstract}
Based on the entropy relations, we derive thermodynamic bound for entropy and area of horizons of Schwarzschild-dS black hole, including the event horizon, Cauchy horizon and negative horizon (i.e. the horizon with negative value), which are all geometrical bound and made up of the cosmological radius. Consider the first derivative of entropy relations together, we get the first law of thermodynamics for all horizons. We also obtain the Smarr relation of horizons by using the scaling discussion. For thermodynamics of all horizons, the cosmological constant is treated as a thermodynamical variable. Especially for thermodynamics of negative horizon, it is defined well in the $r<0$ side of spacetime. The validity of this formula seems to work well for three-horizons black holes. We also generalize the discussion to thermodynamics for event horizon and Cauchy horizon of Gauss-Bonnet charged flat black holes, as the Gauss-Bonnet coupling constant is also considered as thermodynamical variable. These
give further clue on the crucial role that the entropy relations of multi-horizons play in black hole thermodynamics and understanding the entropy at the microscopic level.


\end{abstract}

\section{Introduction}
In order to understand the black hole entropy at the microscopic level, the entropy product of multi-horizons black holes were studied widely in a lot literatures \cite{Cvetic:2010mn,Toldo:2012ec,Cvetic:2013eda,Lu:2013ura,Chow:2013tia,
Detournay:2012ug,Castro:2012av,Visser:2012zi,Chen:2012mh,Castro:2013kea,Visser:2012wu,Abdolrahimi:2013cza,Pradhan:2013hqa,Pradhan:2013xha,
Castro:2013pqa,Faraoni:2012je,Lu:2013eoa,Anacleto:2013esa,Xu:2014qaa,Wang:2013nvz,Wang:2013smb,Xu:2013zpa,Du:2014kpa,Xu:2014qza}.
The entropy product is always independent on the mass of black holes, which is universal for many charged and rotating black holes \cite{Cvetic:2010mn,Toldo:2012ec,Chen:2012mh,Visser:2012zi,Cvetic:2013eda,Abdolrahimi:2013cza,Lu:2013ura,
Anacleto:2013esa,Chow:2013tia,Castro:2013kea,Lu:2013eoa,Wang:2013smb,Xu:2013zpa}, black rings and black strings \cite{Castro:2012av}. Actually, the entropy product, in conjunction with Cauchy horizon thermodynamics, can be used to determine whether the corresponding Bekenstein-Hawking entropy can be written as a Cardy formula, hence providing some evidence for a CFT description of the corresponding microstates \cite{Detournay:2012ug,Castro:2012av}. This makes it important to study the thermodynamics of Cauchy horizon.

On the other hand, the mass-independence of entropy product fails for some multi-horizons black holes \cite{Faraoni:2012je,Castro:2013pqa,Detournay:2012ug,Visser:2012wu,Xu:2014qaa}. Hence, the entropy sum \cite{Xu:2014qaa,Wang:2013nvz,Wang:2013smb,Xu:2013zpa,Du:2014kpa} and other thermodynamic relations \cite{Xu:2014qaa,Visser:2012wu,Pradhan:2013hqa,Pradhan:2013xha,Wang:2013nvz,Xu:2014qza} are introduced, which also have mass-independence for some cases and seem to be universal as well. Especially for the relation $T_{+}S_{+}=T_{-}S_{-}$, which were linked closely with the mass-independence of entropy product. It was also understood well and physically by the holographic description, i.e. the thermodynamic method of black hole/CFT (BH/CFT) correspondence \cite{Chen:2012mh,Chen:2012yd,Chen:2012ps,Chen:2012pt,Chen:2013rb,Chen:2013aza,Chen:2013qza}. The thermodynamic relations $T_{+}S_{+}=T_{-}S_{-}$ may be taken as the criterion whether there is a 2 dimensional CFT dual for the black holes in the Einstein gravity and other diffeomorphism invariant gravity theories \cite{Chen:2012mh,Chen:2012yd,Chen:2012ps,Chen:2012pt,Chen:2013rb,Chen:2013aza,Chen:2013qza}. Besides, It is found that the thermodynamic relation $T_{+}S_{+}=T_{-}S_{-}$ is equivalent to the central charge being the same (i.e. $c_{R} = c_{L}$) for some two-horizons black holes. However, the interpretation of other thermodynamic relations are not clear.
The aim of this work focuses on entropy relations and the application in black hole thermodynamics.

In this paper, based on the entropy relations, we derive thermodynamic bound for entropy and area of horizons of Schwarzschild-dS black hole, including the event horizon, Cauchy horizon and negative horizon, which are all geometrical bound and made up of the cosmological radius. Consider the first derivative of entropy relations together, we get the first law of thermodynamics for all horizons. We also obtain the Smarr relation of horizons by using the scaling discussion. For thermodynamics of all horizons, the cosmological constant is treated as a thermodynamical variable \cite{Kastor:2009wy,Dolan:2011xt,Cvetic:2010jb,Dolan:2013ft,Xu:2013zea,Xu:2014tja,Xu:2014kwa,Altamirano:2014tva}). Especially for thermodynamics of negative horizon, it is also defined well in the negative side ($r<0$). Actually, for black hole solution, there is a singularity, for example located in $r=0$, we always choose the $r>0$ side and the existence of black hole horizons avoids the bare singularity. This makes thermodynamics of positive horizons defined well. Actually, thermodynamics of the event horizon \cite{Lu:2013ura,Chen:2012mh,Abdolrahimi:2013cza,Lu:2013eoa,Detournay:2012ug,Xu:2014qza,Castro:2013pqa,Pradhan:2013hqa,Chen:2012yd,Chen:2012ps,Chen:2012pt,Chen:2013rb,Chen:2013qza}, Cauchy horizon \cite{Lu:2013ura,Chen:2012mh,Abdolrahimi:2013cza,Lu:2013eoa,Castro:2012av,Detournay:2012ug,Xu:2014qza,Castro:2013pqa,Pradhan:2013hqa,Pradhan:2013xha,Chen:2012yd,Chen:2012ps,Chen:2012pt,Chen:2013rb,Chen:2013qza} and cosmological horizon \cite{Barnich:2012xq,Bagchi:2012xr,Riegler:2014bia,Fareghbal:2014qga} are studied widely. For the negative horizon, one can choose the $r<0$ side, and the existence of negative horizons also avoids the bare singularity. On the other hand, it is found that the mass-independence of entropy relations may always hold only when the effect of negative horizon are considered \cite{Wang:2013smb,Xu:2013zpa,Xu:2014qaa,Visser:2012wu,Wang:2013nvz,Du:2014kpa,Xu:2014qza}. This makes it also interesting to study the thermodynamics of negative horizon, even it remains unclear what quantum mechanical degrees of freedom the entropy of the negative horizon count (Note that of the Cauchy horizon is not clear as well). The validity of this formula seems to work well for three-horizons black holes.
We also generalize the discussion to thermodynamics for event horizon and Cauchy horizon of Gauss-Bonnet charged flat black holes, as the Gauss-Bonnet coupling constant is also considered as thermodynamical variable \cite{Kastor:2010gq,Cai:2013qga,Xu:2013zea,Xu:2014tja,Xu:2014kwa,Altamirano:2014tva}). These give further clue on the crucial role that the entropy relations of multi-horizons play in black hole thermodynamics and understanding the entropy at the microscopic level.

This paper is organized as follows. In the next Section, we will investigate the entropy relations and the application of Schwarzschild-dS black hole. In Section 3, the entropy relations and the application of Gauss-Bonnet charged flat black hole are presented. Section 4 is devoted to the conclusions and discussions.

\section{Entropy relations and the application of Schwarzschild-dS black hole}
In this section, we firstly consider the entropy relations and the application of four dimensional  Schwarzschild-dS black hole, which is the simplest example for multi-horizons (A)dS black hole and has the line element
\begin{align}
\mathrm{d}s^2=-f(r)\mathrm{d}t^2+\frac{\mathrm{d}r^2}{f(r)}+r^{2}\left(\mathrm{d}\theta^{2}+\sin^{2}\mathrm{d}\varphi^{2}\right),
\label{SdSMetric}
\end{align}
with the horizon function
\begin{align}
f(r)=1-\frac{2M}{r}-\frac{\Lambda r^2}{3},
\label{SdSf}
\end{align}
where $M$ represents the mass of the black hole and $\Lambda=\frac{1}{\ell^2}$ is the cosmological constant. From the roots of horizon function $f(r)$, we can get three black hole horizons \cite{Visser:2012wu}
\begin{align*}
r_{E} &=2\ell\sin\left(\frac{1}{3} \arcsin \left(\frac{3 M}{\ell}\right)\right)\nonumber\\
r_{C} &=2\ell\sin\left(\frac{1}{3} \arcsin \left(\frac{3 M}{\ell}\right)+\frac{2\pi}{3}\right)\nonumber\\
r_{N} &=2\ell\sin\left(\frac{1}{3} \arcsin \left(\frac{3 M}{\ell}\right)-\frac{2\pi}{3}\right),
\end{align*}
where $r_{E}$, $r_{C}$ and $r_{N}$ denote the event horizon, cosmological horizon and negative horizon, respectively. Note that $r_{N}$ is negative and named as ``virtual'' horizon \cite{Visser:2012wu}. Besides, as we focus on the black hole with multi-horizons, we will only consider the case with
\begin{align}
  \frac{3 M}{\ell}\leq1,
\end{align}
which can be seen as the existence condition of multi-horizons black holes. The entropy of each horizon are
\begin{align}
S_{i}=\frac{A_{i}}{4}=\pi r_i^2,\quad\, (i=E,C,N).
\label{SdSEntropy}
\end{align}
The temperature of event horizon and negative horizon are
\begin{align}
T_{i}=\frac{f^{\prime}(r_{i})}{4\pi}=\frac{\ell^2-r_{i}^2}{4\pi \ell^2r_{i}},\ (i=E,N).
\label{SdSTemE}
\end{align}
while the Hawking temperature of cosmological horizon is always chosen as the positive one \cite{Chen:2012mh}
\begin{align}
  T_{C}=-T_{E}|_{r_{E}\leftrightarrow\,r_{C}}=\frac{r_{C}^2-\ell^2}{4\pi \ell^2r_{C}},
\label{SdSTemC}
\end{align}
where $f^{\prime}(r)$ denotes the derivative function of $f(r)$ respect to $r$.

We firstly revisit the thermodynamic relations. For example, the mass-dependence entropy product is \cite{Visser:2012wu}
\begin{align}
  S_{E}S_{C}S_{N}=\frac{36\pi^3M^2}{\Lambda^2}=36\pi^3M^2\ell^4;\label{SdSProduct}
\end{align}
the mass-independence ``part'' entropy product is \cite{Xu:2014qaa}
\begin{align}
  S_{E}S_{C}+S_{E}S_{N}+S_{C}S_{N}=\frac{9\pi^2}{\Lambda^2}=9\pi^2\ell^4;\label{SdSPart}
\end{align}
entropy sum is \cite{Wang:2013smb,Xu:2014qaa}
\begin{align}
  S_{E}+S_{C}+S_{N}=\frac{6 \pi}{ \Lambda}=6 \pi \ell^2 \label{SdSSum};
\end{align}
and the mass-independent entropy relations of two positive horizons is \cite{Visser:2012wu,Xu:2014qaa}
\begin{align}
  S_{E}+S_{C}+\sqrt{S_{E}S_{C}}=3\pi \ell^2.\label{SdSTwo}
\end{align}

Based on these entropy relations, we can obtain the thermodynamic bound for Schwarzschild-dS black hole. As $0\leq\,r_{E}\leq\,r_{C}\leq|r_{N}|\leq2\ell$, we get $0\leq\,S_{E}\leq\,S_{C}\leq\,S_{N}\leq4\pi\ell^2$.
Thus, from thermodynamic relation Eq.(\ref{SdSTwo}), we get
\begin{align*}
  0\leq3S_{E}\leq(S_{E}+S_{C}+\sqrt{S_{E}S_{C}})=3\pi\ell^2\leq3\,S_{C},
\end{align*}
and
\begin{align*}
  0\leq\,S_{C}\leq3\pi\ell^2,
\end{align*}
which together give
\begin{align*}
  0\leq\,S_{E}\leq\pi\ell^2\leq\,S_{C}\leq3\pi\ell^2.
\end{align*}
Meanwhile, thermodynamic relation Eq.(\ref{SdSTwo}) also leads to $0\leq\,(S_{C}+S_{E})\leq3\pi\ell^2$; hence, from the entropy sum Eq.(\ref{SdSSum}), we find
\begin{align*}
  \,S_{N}\geq3\pi\ell^2.
\end{align*}
Totally, we obtain the entropy bound of the event horizon, the cosmological horizon and the negative horizon
\begin{align}
\,S_{E}\in\bigg[0,\pi\ell^2\bigg],\quad\,S_{C}\in\bigg[\pi\ell^2,3\pi\ell^2\bigg],\quad\,S_{N}\in\bigg[3\pi\ell^2,4\pi\ell^2\bigg].
\end{align}
Besides, the area entropy leads to the area bound
\begin{align}
\sqrt{\frac{\,A_{E}}{16\pi}}\in\bigg[0,\frac{\ell}{2}\bigg],\quad\,\sqrt{\frac{\,A_{C}}{16\pi}}\in\bigg[\frac{\ell}{2},\sqrt{\frac{3}{4}}\ell\bigg],\quad\,\sqrt{\frac{\,A_{N}}{16\pi}}\in\bigg[\sqrt{\frac{3}{4}}\ell,\ell\bigg],
\end{align}
which are all geometrical bounds of black hole horizons, as parameter $\ell$ is actually the cosmological radius.

On the other hand, we can get the first law of thermodynamics from these thermodynamic relations.
Actually, thermodynamics of (A)dS black holes are  still open questions. An interesting idea is treating the cosmological constant as a thermodynamical variable (see, e.g. \cite{Kastor:2009wy,Dolan:2011xt,Cvetic:2010jb,Dolan:2013ft,Xu:2013zea,Xu:2014tja,Xu:2014kwa,Altamirano:2014tva}). Hence, consider the first derivative of thermodynamic relations Eq.(\ref{SdSProduct}, \ref{SdSPart}, \ref{SdSSum}), one can find
\begin{align*}
  &S_{C}S_{N}\mathrm{d}S_{E}+S_{E}S_{C}\mathrm{d}S_{N}+S_{N}S_{E}\mathrm{d}S_{C}=72\pi^3\bigg(\frac{M}{\Lambda^2}\mathrm{d}M-\frac{M^2}{\Lambda^3}\mathrm{d}\Lambda\bigg),\\
  &(S_{C}\mathrm{d}S_{E}+S_{E}\mathrm{d}S_{C})+(S_{C}\mathrm{d}S_{N}+S_{N}\mathrm{d}S_{C})+(S_{E}\mathrm{d}S_{N}+S_{N}\mathrm{d}S_{E})=-\frac{18\pi^2}{\Lambda^3}\mathrm{d}\Lambda,\\
  &\mathrm{d}S_{E}+\mathrm{d}S_{C}+\mathrm{d}S_{N}=-\frac{6\pi}{\Lambda^2}\mathrm{d}\Lambda.
\end{align*}
These lead to
\begin{align*}
  &\mathrm{d}S_{E}=-\frac{72\pi^3M}{(S_{E}-S_{N})(S_{C}-S_{E})\Lambda^2}\mathrm{d}M+\frac{6\pi(12\pi^2M^2-3\pi\,S_{E}+\Lambda\,S_{E}^2)}{(S_{E}-S_{N})(S_{C}-S_{E})\Lambda^3}\mathrm{d}\Lambda,\\
  &\mathrm{d}S_{C}=\frac{72\pi^3M}{(S_{C}-S_{N})(S_{C}-S_{E})\Lambda^2}\mathrm{d}M-\frac{6\pi(12\pi^2M^2-3\pi\,S_{C}+\Lambda\,S_{C}^2)}{(S_{C}-S_{N})(S_{C}-S_{E})\Lambda^3}\mathrm{d}\Lambda,\\
  &\mathrm{d}S_{N}=\frac{72\pi^3M}{(S_{E}-S_{N})(S_{C}-S_{N})\Lambda^2}\mathrm{d}M-\frac{6\pi(12\pi^2M^2-3\pi\,S_{N}+\Lambda\,S_{N}^2)}{(S_{E}-S_{N})(S_{C}-S_{N})\Lambda^3}\mathrm{d}\Lambda.
\end{align*}
or equivalently
\begin{align*}
  &\mathrm{d}M=\frac{\Lambda^2(S_{E}-S_{N})(S_{E}-S_{C})}{72\pi^3M}\mathrm{d}S_{E}+\frac{(12\pi^2M^2-3\pi\,S_{E}+\Lambda\,S_{E}^2)}{12\pi^2M\Lambda}\mathrm{d}\Lambda,\\
  &\mathrm{d}M=-\frac{\Lambda^2(S_{C}-S_{N})(S_{E}-S_{C})}{72\pi^3M}\mathrm{d}S_{C}+\frac{(12\pi^2M^2-3\pi\,S_{C}+\Lambda\,S_{C}^2)}{12\pi^2M\Lambda}\mathrm{d}\Lambda,\\
  &\mathrm{d}M=\frac{\Lambda^2(S_{C}-S_{N})(S_{E}-S_{N})}{72\pi^3M}\mathrm{d}S_{N}+\frac{(12\pi^2M^2-3\pi\,S_{N}+\Lambda\,S_{N}^2)}{12\pi^2M\Lambda}\mathrm{d}\Lambda.
\end{align*}
Consider the Hawking temperature (\ref{SdSTemE},\ref{SdSTemC}), we get the first law of thermodynamics for event horizon, cosmological horizon and negative horizon of Schwarzschild-dS black hole
\begin{align}
  &\mathrm{d} M=+\,T_{E}\mathrm{d} S_{E}+V_{E} \mathrm{d} \Lambda,\label{SdSFirstLawE}\\
  &\mathrm{d} M=-\,T_{C}\mathrm{d} S_{C}+V_{C} \mathrm{d} \Lambda,\label{SdSFirstLawC}\\
  &\mathrm{d} M=-\,T_{N}\mathrm{d} S_{N}+V_{N} \mathrm{d} \Lambda.\label{SdSFirstLawN}
\end{align}
where the thermodynamic potential conjugate to $\Lambda$ is defined to be
\begin{align}
  V_{i}=\bigg(\frac{\partial M}{\partial \Lambda}\bigg)_{S_{i}}=-\frac{r_{i}^3}{6}=\frac{(12\pi^2M^2-3\pi\,S_{i}+\Lambda\,S_{i}^2)}{12\pi^2M\Lambda},\quad (i=E,C,N).
\end{align}

Furthermore, the Smarr relation for horizons can be found by scaling arguments. The mass can be viewed as a homogeneous function of the thermodynamical variables $S_{i}$ and $\Lambda$, i.e. $M=M(S_{i}, \Lambda)$. From the horizon function Eq.(\ref{SdSf}), one can find that mass $M$ scales as [length]$^1$ and $\Lambda$ scales as [length]$^{-2}$. The area entropy Eq.(\ref{SdSEntropy}) shows that $S_{i}$ scales as [length]$^2$. Then after a rescaling of the thermodynamical variables, we can get $
\lambda^{1} M =M(\lambda^{2} S_{i}, \lambda^{2} \Lambda). $
Taking the first derivative with respect to $\lambda$ and then setting
$\lambda=1$, we get the Smarr relation for the event horizon and negative horizon
\begin{align}
&M =2\bigg( T_{E} S_{E}+V_{E}\Lambda\bigg),
\label{SdSSmarrE}\\
&M =2\bigg( T_{N} S_{N}+V_{N}\Lambda\bigg).
\label{SdSSmarrN}
\end{align}
Note we choose the positive temperature Eq.(\ref{SdSTemC}), other than the origin negative (opposite) one, thus the Smarr relation for the cosmological horizon of Schwarzschild-dS black hole is
\begin{align}
M=2\bigg(-\,T_{C} S_{C}+V_{C}\Lambda\bigg).
\label{SdSSmarrC}
\end{align}
Finally, we obtain the first law of thermodynamics Eq.(\ref{SdSFirstLawE}, \ref{SdSFirstLawC}, \ref{SdSFirstLawN}) and Smarr relation Eq.(\ref{SdSSmarrE}, \ref{SdSSmarrC}, \ref{SdSSmarrN}) for the event horizon, the cosmological horizon and negative horizon of Schwarzschild-dS black hole.

For Schwarzschild-dS black hole with $\frac{3 M}{\ell}>1$, one can only find one real root of $f(r)$ which is the event horizon and the other two are complex. This case is out of our discussion for the reason that we study the thermodynamic laws of horizons in this paper, while the thermodynamics of complex horizon are not defined well. On the other hand, the four dimensional uncharged black hole in $f(R)$ gravity \cite{Castro:2013pqa} has the same line element Eq.(\ref{SdSMetric}) with different metric function $f(r)=1-\frac{2\mu}{r}-\frac{R_{0}}{12}r^2$, where $R=R_0$ is the constant curvature of the static, spherically symmetric solution. Hence, following the same procedure, we can obtain similar thermodynamic relations, thermodynamic bound of entropy and area, first law of thermodynamics and Smarr relations, with the cosmological constant being $\Lambda_{f}=\frac{R_0}{4}$ and the cosmological radius being $\ell_{f}=\frac{2}{\sqrt{R_0}}$. Actually, one can always expect to follow the similar procedure to get these results of other black holes with three horizons; for example, four dimensional charged static black holes in Einstein-Weyl theory \cite{Cvetic:2013eda} and five dimensional charged (A)dS black holes in the Gauss-Bonnet gravity \cite{Castro:2013pqa}.

\section{Entropy relations and the application of A Gauss-Bonnet black hole example}
In this section, we introduce a Gauss-Bonnet black hole example to study further the thermodynamic relations and the application. However, we will only consider the positive horizons for this case, which had attracted much attentions \cite{Lu:2013ura,Chen:2012mh,Abdolrahimi:2013cza,Lu:2013eoa,Castro:2012av,Detournay:2012ug,Xu:2014qza,Castro:2013pqa,Pradhan:2013hqa,Pradhan:2013xha,Chen:2012yd,Chen:2012ps,Chen:2012pt,Chen:2013rb,Chen:2013qza}, without the negative ones. We consider the five dimensional charged asymptotically flat solutions. We take the action to be
\begin{align*}
  S=\frac{1}{16\pi\,G}\int\mathrm{d}^5x\sqrt{-g}(\mathcal{L}_{0}+\mathcal{L}_{1}+\mathcal{L}_{GB})+\int\mathrm{d}^5x\sqrt{-g}\mathcal{L}_{matter},
\end{align*}
where
\begin{align*}
  &\mathcal{L}_{0}=-2\Lambda=0,\quad\mathcal{L}_{1}=R,\quad\mathcal{L}_{matter}=\frac{1}{2}F_{\mu\nu}F^{\mu\nu},\\
  &\mathcal{L}_{GB}=\alpha(R_{\mu\nu\lambda\rho}R^{\mu\nu\lambda\rho}-4R_{\mu\nu}R^{\mu\nu}+R^2)
\end{align*}
Here we have chosen the vanishing cosmological constant and rescaled the Gauss-Bonnet coupling constant $\tilde{\alpha}$ in the following discussion by
\begin{align*}
  \tilde{\alpha}=(d-3)(d-4)\alpha=2\alpha.
\end{align*}
The static and charged black hole solution has the form \cite{Boulware:1985wk,Wheeler:1985nh,Cai:2001dz,Wiltshire:1985us,Cai:2003kt}
\begin{align*}
  \mathrm{d}s^2=-V(r)\mathrm{d}t^2+\frac{\mathrm{d}r^2}{V(r)}+r^2\mathrm{d}\Omega_{3}^2,\quad\,F=\frac{q}{4\pi\,r^3}\mathrm{d}t\wedge\mathrm{d}r,
\end{align*}
where $\mathrm{d}\Omega_{3}^2$ is the maximally symmetric space in $3$-dimensions and the metric function is
\begin{align}
  V(r)=1+\frac{r^2}{2\tilde{\alpha}}\bigg[1-\sqrt{1+4\tilde{\alpha}\bigg(\frac{2\mu}{r^4}-\frac{q^2}{r^6}\bigg)}\bigg].
\label{GBV}
\end{align}
The event horizon $r_{E}$ and Cauchy horizon $r_{C}$ are located in the roots of the metric function $V(r)$
\begin{align}
  &r_{E}^2=\frac{1}{2}(2\mu-\tilde{\alpha})+\frac{1}{2}\sqrt{(2\mu-\tilde{\alpha})^2-4q^2},\\
  &r_{C}^2=\frac{1}{2}(2\mu-\tilde{\alpha})-\frac{1}{2}\sqrt{(2\mu-\tilde{\alpha})^2-4q^2}.
\end{align}
Note we will only consider the positive horizons in what follows.
We introduce some useful relations: $r_{E}r_{C}=q, r_{E}+r_{C}=\sqrt{2\mu-\tilde{\alpha}+2q}, r_{E}^2+r_{C}^2=2\mu-\tilde{\alpha}$.  The temperatures, electric potentials, areas and entropies for horizons are given by
\begin{align}
  &T_{E}=\frac{r_{E}^2-r_{C}^2}{2\pi\,r_{E}(2\tilde{\alpha}+r_{E}^2)},\quad T_{C}=\frac{r_{E}^2-r_{C}^2}{2\pi\,r_{C}(2\tilde{\alpha}+r_{C}^2)}\label{GBTemC}\\
  &\Phi_{E}=2\bigg(\frac{\pi}{4}\bigg)^{1/3}\frac{Q}{r_{E}^2},\quad\quad\,\Phi_{C}=2\bigg(\frac{\pi}{4}\bigg)^{1/3}\frac{Q}{r_{C}^2}\\
  &A_{E}=2\pi^2r_{E}^3,\quad\quad\quad\quad\quad\,A_{C}=2\pi^2r_{C}^3\\
  &S_{E}=\frac{\pi^2r_{E}^3}{2}\bigg(1+\frac{6\tilde{\alpha}}{r_{E}^2}\bigg),\quad\,S_{C}=\frac{\pi^2r_{C}^3}{2}\bigg(1+\frac{6\tilde{\alpha}}{r_{C}^2}\bigg),\label{GBS0}
\end{align}
where the ADM mass $M$ and the electric charge $Q$ of the solution are given by
\begin{align}
  M=\frac{3\pi\mu}{4},\quad\,Q=\left(\frac{\pi}{4}\right)^{2/3}q.
\end{align}
We can find the entropy product \cite{Castro:2013pqa} and entropy sum
\begin{align}
  S_{E}S_{C}=4\pi^2\bigg(1+\frac{12\tilde{\alpha}\mu}{q^2}+\frac{30\tilde{\alpha}^2}{q^2}\bigg)Q^3,\quad\,S_{E}+S_{C}=\frac{\pi^2}{2}\sqrt{2\mu-\tilde{\alpha}+2q}\bigg(2\mu+5\tilde{\alpha}-q\bigg),
\label{GBS}
\end{align}
and the area product \cite{Castro:2013pqa} and area sum
\begin{align}
  A_{E}A_{C}=64\pi^2Q^3,\quad\,A_{E}+A_{C}=2\pi^2\sqrt{2\mu-\tilde{\alpha}+2q}(2\mu-\tilde{\alpha}-q).
\end{align}

The existence of black hole horizons leads to
\begin{align}
  \mu\geq\,q+\frac{\tilde{\alpha}}{2}.
\end{align}
Focus on the area bound of horizons, we get
\begin{align*}
  A_{E}\geq\sqrt{A_{E}A_{C}}=9\pi\,Q\sqrt{Q},\quad\,A_{C}\leq\sqrt{A_{E}A_{C}}=9\pi\,Q\sqrt{Q}
\end{align*}
The area sum gives
\begin{align*}
  &\pi^2\sqrt{2\mu-\tilde{\alpha}+2q}(2\mu-\tilde{\alpha}-q)=\frac{A_{E}+A_{C}}{2}\leq\,A_{E}\leq\,A_{E}+A_{C}=2\pi^2\sqrt{2\mu-\tilde{\alpha}+2q}(2\mu-\tilde{\alpha}-q),\\
  &A_{C}\leq\frac{A_{E}+A_{C}}{2}=\pi^2\sqrt{2\mu-\tilde{\alpha}+2q}(2\mu-\tilde{\alpha}-q).
\end{align*}
Consider the above bound and the existence of black hole horizons together, we obtain the area bound of event horizon and Cauchy horizon
\begin{align}
  A_{E}\in\bigg[\pi^2,2\pi^2\bigg]\times\sqrt{2\mu-\tilde{\alpha}+2q}(2\mu-\tilde{\alpha}-q),\quad\,A_{C}\in\bigg[0,9\pi\,Q\sqrt{Q}\bigg].
\end{align}
One can also obtain the entropy bound as well, which is complicated and does not shown here.

On the other hand, consider the first derivative of entropy product and sum Eq.(\ref{GBS}) and make a little calculation which is similar as that in the above section, one can find the first law of thermodynamics for event horizon and Cauchy horizon of Gauss-Bonnet charged flat black hole
\begin{align}
  &\mathrm{d} M=+\,T_{E}\mathrm{d} S_{E}+\Phi_{E} \mathrm{d} Q+\Theta_{E}\mathrm{d}\tilde{\alpha},\label{GBFirstLawE}\\
  &\mathrm{d} M=-\,T_{C}\mathrm{d} S_{C}+\Phi_{C} \mathrm{d} Q+\Theta_{C}\mathrm{d}\tilde{\alpha}.\label{GBFirstLawC}
\end{align}
Note that here we have treated the Gauss-Bonnet coupling constant as a thermodynamical variable (see, e.g. \cite{Kastor:2010gq,Cai:2013qga,Xu:2013zea,Xu:2014tja,Xu:2014kwa,Altamirano:2014tva}). Accordingly, the thermodynamic potential conjugate to $\tilde{\alpha}$ is defined to be
\begin{align}
  &\Theta_{E}=\bigg(\frac{\partial M}{\partial \tilde{\alpha}}\bigg)_{S_{E},Q}=\frac{3\pi}{8}\bigg(1-8\pi\tilde{\alpha}r_{E}T_{E}\bigg),\\
  &\Theta_{C}=\bigg(\frac{\partial M}{\partial \tilde{\alpha}}\bigg)_{S_{C},Q}=\frac{3\pi}{8}\bigg(1+8\pi\tilde{\alpha}r_{C}T_{C}\bigg).
\end{align}
Besides, the Smarr relation for horizons can be found by scaling arguments. The mass can be viewed as a homogeneous function of the thermodynamical variables $S_{i}$ and $\tilde{\alpha}$, i.e. $M=M(S_{i}, \tilde{\alpha})$. From the horizon function Eq.(\ref{GBV}), one can find that mass $M$ scales as [length]$^2$, electric charge $Q$ scales as [length]$^{2}$ and $\tilde{\alpha}$ scales as [length]$^{2}$. The area entropy Eq.(\ref{GBS0}) shows that $S_{i}$ scales as [length]$^3$. Then after a rescaling of the thermodynamical variables, we can get $
\lambda^{2} M =M(\lambda^{3} S_{i}, \lambda^{2} Q, \lambda^{2} \tilde{\alpha}). $
Taking the first derivative with respect to $\lambda$ and then setting
$\lambda=1$, we get the Smarr relation for the event horizon
\begin{align}
&M =+\frac{3}{2} T_{E} S_{E}+\Phi_{E}Q+\Theta_{E}\tilde{\alpha},
\label{GBSmarrE}
\end{align}
Note we choose the positive temperature Eq.(\ref{GBTemC}), other than the origin negative (opposite) one, thus the Smarr relation for the Cauchy horizon is
\begin{align}
&M =-\frac{3}{2} T_{C} S_{C}+\Phi_{C}Q+\Theta_{C}\tilde{\alpha}.
\label{GBSmarrC}
\end{align}
Finally, we obtain the first law of thermodynamics Eq.(\ref{GBFirstLawE}, \ref{GBFirstLawC}) and Smarr relation Eq.(\ref{GBSmarrE}, \ref{GBSmarrC}) for the event horizon and the Cauchy horizon of Gauss-Bonnet charged flat black hole, which is consistence with that in \cite{Castro:2013pqa}.

\section{Conclusions}
In this paper, based on the entropy relations, we obtain the thermodynamic bound of entropy and areafor horizons of Schwarzschild-dS black hole, including the event horizon, Cauchy horizon and negative horizon, which are all geometrical bound and made up of the cosmological radius. Consider the first derivative of entropy relations together, we get the first law of thermodynamics for all horizons. We also obtain the Smarr relation of horizons by using the scaling discussion. For thermodynamics of all horizons, the cosmological constant is treated as a thermodynamical variable. Especially for thermodynamics of negative horizon, it is also defined well in the negative side ($r<0$).  The validity of this formula seems to work well for three-horizons black holes; for example, four dimensional uncharged black hole in $f(R)$ gravity \cite{Castro:2013pqa}, four dimensional charged static black holes in Einstein-Weyl theory \cite{Cvetic:2013eda} and five dimensional charged (A)dS black holes in the Gauss-Bonnet gravity \cite{Castro:2013pqa}.
We also generalize the discussion to thermodynamics for event horizon and Cauchy horizon of Gauss-Bonnet charged flat black holes, as the Gauss-Bonnet coupling constant is also considered as thermodynamical variable. These give further clue on the crucial role that the entropy relations of multi-horizons play in black hole thermodynamics and understanding the entropy at the microscopic level.

For future work, one can believe that the validity of this formula holds for general Lovelock gravity hence more coupling constant entering in the laws of black hole thermodynamics. Besides, as the thermodynamics of the negative horizon is introduced, one may also expect to construct the holographic description of thermodynamics for black holes with three-horizons, while the holographic description of thermodynamics for black holes with two-horizons are well built by the thermodynamic method of black hole/CFT (BH/CFT) correspondence \cite{Chen:2012mh,Chen:2012yd,Chen:2012ps,Chen:2012pt,Chen:2013rb,Chen:2013aza,Chen:2013qza}.

\section*{Acknowledgments}
We would like to thank professor M. Cvetic, Jian-wei Mei, C.N. Pope and Liu Zhao for useful conversations. This work is partially supported by the Natural Science Foundation of China (NSFC) under Grant No.11075078. Wei Xu was supported by the Research Innovation Fund of Huazhong University of Science and Technology (2014TS125 and 0128012029).

\providecommand{\href}[2]{#2}\begingroup
\footnotesize\itemsep=0pt
\providecommand{\eprint}[2][]{\href{http://arxiv.org/abs/#2}{arXiv:#2}}

\end{document}